# Expanded microchannel heat exchanger: design, fabrication and preliminary experimental test

**David C. Denkenberger**[a*], **Michael J. Brandemuehl**[b], **Joshua M. Pearce**[c], **and John Zhai**[b]

[a]Denkenberger Inventing and Consulting, LLC, 2345 Forest Ave, Durango CO 81301;
David.Denkenberger@colorado.edu, 1.970.259.6801x322 (p), 1. 970.259.8585 (f)
[b]Civil, Environmental, and Architectural Engineering Department, University of Colorado at Boulder, 428 UCB, Boulder, CO 80309, USA, Michael.Brandemuehl@colorado.edu, John.Zhai@colorado.edu
[c]Department of Materials Science & Engineering and Department of Electrical & Computer Engineering Michigan Technological University, 601 M&M Building, 1400 Townsend Drive, Houghton, MI 49931, pearce@mtu.edu

**Abstract**
This paper first reviews non-traditional heat exchanger geometry, laser welding, practical issues with microchannel heat exchangers, and high effectiveness heat exchangers. Existing microchannel heat exchangers have low material costs, but high manufacturing costs. This paper presents a new expanded microchannel heat exchanger design and accompanying continuous manufacturing technique for potential low-cost production. Polymer heat exchangers have the potential for high effectiveness. The paper discusses one possible joining method - a new type of laser welding named "forward conduction welding," used to fabricate the prototype. The expanded heat exchanger has the potential to have counter-flow, cross-flow, or parallel-flow configurations, be used for all types of fluids, and be made of polymers, metals, or polymer-ceramic precursors. The cost and ineffectiveness reduction may be an order of magnitude or more, saving a large fraction of primary energy. The measured effectiveness of the prototype with 28 micron thick black low density polyethylene walls and counterflow, water-to-water heat transfer in 2 mm channels was 72%, but multiple low-cost stages could realize the potential of higher effectiveness.

**1. Introduction**

Heat exchangers (HXs) transfer heat from one fluid to another (both liquids and gases are considered fluids). HXs are used in refrigeration cycles, heat recovery, industrial processes, vehicles, and conventional power plants [1]. Also, renewable energy applications include fuel cells, concentrated solar power, solar hot water, compressed air energy storage [2], wind turbines, geothermal, and solar water pasteurization. The latter application has the potential to save the lives of thousands of children a day [3], [4]. The performance of a HX is characterized by its effectiveness, $\eta$, defined as the actual heat transfer rate as a fraction of the maximum heat transfer rate:

$$\eta = \frac{\dot{q}}{\dot{q}_{max}}, \qquad \textbf{Equation 1}$$

where the actual heat transfer rate is:

$$\dot{q} = C_h(T_{hi} - T_{ho}) = C_c(T_{co} - T_{ci}), \qquad \textbf{Equation 2}$$

where $T_{hi}$ is the temperature of the hot fluid going into the HX, $T_{ho}$ is the temperature of the hot fluid coming out of the HX, $T_{ci}$ is the temperature of the cold fluid going into the HX, and $T_{co}$ is the temperature of the cold fluid coming out of the HX [5]. Also, the hot heat capacity rate is:

$$C_h = \dot{m}_h C_{ph}, \qquad \textbf{Equation 3}$$

where $\dot{m}_h$ is the hot mass flow rate and $C_{ph}$ is the hot specific heat. Furthermore, the cold heat capacity rate is:

$$C_c = \dot{m}_c C_{pc}, \qquad \textbf{Equation 4}$$

where $\dot{m}_c$ is the cold mass flow rate and $C_{pc}$ is the cold specific heat. The maximum heat transfer rate is:

$$\dot{q}_{max} = C_{min}(T_{hi} - T_{ci}), \qquad \textbf{Equation 5}$$

where $C_{min}$ is the minimum of $C_h$ and $C_c$.

For a counterflow HX, the *NTU* model is

$$\eta = \frac{1 - e^{-NTU(1-C)}}{1 - Ce^{-NTU(1-C)}} \qquad \textbf{Equation 6}$$

where $C$ is heat capacity flow ratio = $C_{min}/C_{max}$; NTU is "number of transfer units:"

$$NTU = \frac{hA}{C_{min}}, \qquad \text{Equation 7}$$

where $A$ is the heat transfer area and $h$ is the heat transfer coefficient [5].
When $C = 1$, the HX is said to have balanced flow. For $C=1$,

$$\eta = \frac{NTU}{1+NTU}, \qquad \text{Equation 8}$$

[6]. For high effectiveness HXs, it is more convenient to work with the ineffectiveness, $(1-\eta)$, so for $C = 1$,

$$(1 - \eta) = \frac{1}{1+NTU}. \qquad \text{Equation 9}$$

Significantly increasing the effectiveness of HXs would save a large fraction of primary energy, which can be achieved with conventional technology assuming energy efficiency is viewed as an investment [7]. This would decrease $CO_2$ and other emissions by a similar amount through fossil energy efficiency.

Microchannel HXs (hydraulic diameter < 1 mm) are currently used, and they have low material cost, weight, and volume, but the manufacturing techniques are expensive. Methods of producing microchannel HXs are etching, LIGA (lithography, electroplating and molding), micromachining, and stereolithography [8].

This paper determines the technical viability of high effectiveness microchannel HXs. First non-traditional HX geometry, laser welding, practical issues with microchannel HXs, high effectiveness HXs, and polymer HXs with potential high effectivenesses are reviewed. Then a novel polymer-based HX design is proposed and manufacturing technique involving laser welding and expansion to produce low-cost microchannel HXs is described. A prototype is constructed using the method and design and tested using a thermal bath technique. Experimental results are discussed, future work is outlined and conclusions are drawn.

## 2. Background

This section reviews non-traditional HX geometry, laser welding, practical issues with microchannel HXs, and high effectiveness HXs. The review will explore the current status, challenges, opportunities, and potential approaches to developing low-cost high effectiveness HXs.

*2.1 Square and Triangular HX channels*

There are a number of methods of producing square (see Figure 1) or triangular HX channels [9-13]. All have the difficulty of aligning the core of the HX, where the majority of heat transfer occurs, with the manifold, which takes the large diameter flow (input and output tubes) and distributes the flow to the small channels in the core. Therefore, these techniques cannot produce microchannels. Carman [11] proposes solidifying a polymer with laser, producing 0.05 mm walls. Then the polymer is pyrolyzed into a ceramic to handle high fluid pressures. Perry [12] proposes welding two plastic layers together, and then stacking them.

Figure 1. Cross-section end view of the extruded "chessboard" pattern [11].

*2.2 Laser Welding*

Laser welding is one way of fabricating HXs. The conventional technique for polymer laser welding requires the upper polymer layer to be transparent to the laser, and the lower layer to be opaque to the laser. This lower layer absorbs the electromagnetic radiation, melts, and melts the transparent layer immediately above it, forming a weld [14].

For welding of polymers, the temperature must be above the melting point, but below the decomposition point. This temperature difference, or viable range, influences how easy it is to laser weld a polymer [14]. The best polymers for this are high density polyethylene (HDPE), polypropylene (PP), and polystyrene (PS).

Another laser welding technique is called reverse conduction laser welding [15]. The laser passes through clear polymer layers, is absorbed by a black metal substrate, and then the heat conducts up through the polymer to weld two polymer layers together. However, if the spot is exposed too long, the heat conducts all the way through two layers and into the third layer and bonds three layers together (see section 4.3 for a similar figure).

*2.3 Failure Mechanisms and Practical Solutions*

Two additional concerns with microchannel HXs are fouling and erosion.

There are variety of solutions to chemical fouling, including dilution, prevention of deposition on the walls by magnetic means, and scale removal by physical or chemical means [16]. Fouling can also be due to physical means, for instance, particle deposition. It is recommended that the particles be less than one third the channel size to avoid clogging [17]. Since fouling with polymers is less of a problem than with metals because the polymers are hydrophobic and expand more with temperature changes, shedding fouling coatings [1], the particles could be at least as large as one third the channel size for polymer walls.

The wall of the HX can also be physically eroded. The erosion rate of polymer pipes is generally less than that of metal pipes, and the wear decreases with particle size and velocity, with a higher exponent than unity [18]. Assuming that the wall thickness scales with the channel dimension, the particle size will scale with the wall thickness to avoid clogging, and since the erosion rate falls faster than the particle size, this means that percent erosion rate of the wall thickness would be smaller. Therefore, wall erosion is not likely to be a serious problem for microchannel HXs, especially polymer ones.

*2.4 Example High effectiveness HXs*

Typical HX effectiveness is 60-80%; however, certain applications benefit from higher values and have promoted the development of high effectiveness HXs. One such application is the destruction of organic compounds, such as chemical weapons, with the flow of hot air through a catalyst. A microchannel HX with 0.8 mm channels achieved $\eta = 97\%$ [19]. Ceramic materials were used to overcome axial conduction (an effectiveness loss mechanism in the same direction as the fluid flow) problems with stainless steel. A fuel cell application has produced $\eta = 97\%$ for gas-to-gas heat exchange [20]. A gas turbine regenerator has achieved $\eta = 98\%$ [21], which helps to achieve an overall micro-turbine cycle efficiency of 50% and rivals that of central power generation.

Cryogenic refrigeration cycles require high effectiveness to even function. A helium HX in a space application has achieved $\eta = 99.8\%$ [22].

**3. Achieving High Effectiveness with Polymer Microchannel HXs**

Doty [6] analyzed the case of microchannel HXs. Most current HXs utilize turbulent flow, which produces high heat transfer coefficient. At smaller channel hydraulic diameter, the flow becomes laminar, which reduces heat transfer coefficient. However, at even smaller hydraulic diameters, the distance that the heat has to conduct through the fluid becomes short, so heat transfer coefficient increases, even exceeding the turbulent heat transfer coefficient for very small channels. Furthermore, with laminar flow, when velocity is decreased, heat transfer coefficient is maintained (constant Nusselt number (Nu), which is the ratio of the thermal conductance in the fluid with convection to the thermal conductance without convection). Since the head loss decreases with lower velocity, the head loss can be made very low by having many parallel channels.

Polymers are already being used for HXs in corrosive environments, such as industrial applications [1], condensing furnaces [23], and have been proposed for ocean thermal energy conversion [24]. Also, the low cost of polymers is attractive in air-to-air applications, where the thermal resistance in the fluid is high [25]. In high pressure applications, where the polymer cannot be made as thin, additives such as graphite can increase the thermal conductivity of the polymer [26]. Also, fiber reinforcement increases strength and temperature resistance of polymers, allowing one HX to be designed to withstand a pressure of 60 atmospheres, which is sufficient for refrigerant cycles [1].

The low thermal conductivity of polymers is a barrier to achieving high effectiveness. However, the limitations can be overcome with proper material thickness and channel size. Consider a polymer microchannel HX with adjacent same-sized tubes, which can be achieved with square passages (see Figure 1) or triangular passages. For non-finned surfaces, no fouling, thin walls, and equal convection coefficients on both sides, the overall heat transfer coefficient is:

$$U = \frac{1}{\frac{2}{h} + \frac{t_w}{k_w}}, \qquad \textbf{Equation 10}$$

where $h$ is the heat transfer coefficient in each of the fluids and $t_w$ and $k_w$ are the thickness and thermal conductivity of the wall, respectively. The effect of wall thickness is shown in Figure 2 for a material with $k_w = 0.2$ W/(mK), typical of polymers. Note that with a typical metal $k_w = 100$ W/(m*K), 1mm thickness has negligible effect. The following conclusions can be drawn for various flow configurations:

- 1 mm polymer wall thickness has little effect when $h < 100$ W/(m$^2$K), which can be achieved with turbulent gas flow or with laminar gas flow in 1 mm diameter channels.

- 0.1 mm polymer wall thickness has little effect when $h<1000$ W/(m$^2$K), which can be achieved with laminar gas flow in 0.1 mm diameter tubes or with turbulent liquid flow or laminar liquid flow in 1 mm channels.
- 0.01 mm polymer wall thickness has little effect when $h<10,000$ W/m$^2$ K, which can be achieved with laminar liquid flow in 0.1 mm diameter channels.

Figure 2. Overall heat transfer coefficient with metal and different polymer thicknesses (mm).

Other factors also reduce the penalty of low polymer thermal conductivity. Surface fouling, on both metal and polymer HXs, increases the thermal resistance and makes the lower conductivity materials relatively less important. Also, the low thermal conductivity of a polymer can actually be an advantage for high effectiveness HXs, because it reduces axial conduction.

A significant benefit of polymer HXs is low material price. Multiplying the thickness of the material $t$ and the price of the material $P_V$ yields the price per area. Dividing this by the overall heat transfer coefficient yields the material price per heat transfer ability:

$$P_{HT} = P_V * \frac{t}{U}.$$  **Equation 11**

The scaling advantage is shown in Figure 3. For a constant ratio between wall thickness and channel diameter (in this case 0.1), both the overall thermal resistance and material price per heat transfer area scale with the diameter (slope = 1 on the log-log plot). The product of these $P_{HT}$ varies with the square of the diameter, or slope = 2. Therefore, a small diameter is highly advantageous.

Figure 3. Price per area ($/m$^2$), thermal resistance (m$^2$/(W/K)), and price per heat transfer ability ($/(W/K)) as a function of channel diameter (mm) with wall thickness 10% of diameter.

The analysis suggests that microchannel polymer HXs can be designed with small channel diameter, thin walls, low velocity, short flow lengths, and large face areas to produce high effectiveness, low initial cost (with potential mass production), and low pumping energy costs.

**4. New Expanded HX**
*4.1 Design*

A design and method of manufacturing inexpensively HXs with small characteristic dimension has been proposed [27]. Sheets are bonded together in strips and pulled apart (see Figure 4). In reality, there would be non-zero radii of curvature (not perfectly angular rectangles). A similar process is currently used for honeycomb manufacture with aluminum, but the aluminum actually stretches upon expansion, requiring very strong welds. For the expansion ("non-stretching expansion") considered here, one could use adhesive or thermal welding (contact or electromagnetic radiation). Low-cost sheets can be made ~10 µm thick (e.g. aluminum foil and plastic wrap). The accuracy of inkjet printing can be 5 µm with a line width of 50 µm [28]. Laser welding can have an accuracy of 2 µm with a line width of 100 µm [29]. Therefore, it appears that channels 200 µm in diameter or smaller could be made with this technique.

Figure 4. Cross section view of the expansion process for the HX core. The start is the top and the bottom is expanded (with insulation put on the outside).

The manifold is made in the same expansion process (see Figure 5). Fewer welds are made near the ends of the sheets. In this way, the "chessboard" pattern in the HX core is converted to fluid sheets of alternating hot and cold (see Figure 6). Then the cross-flow manifold separates the two flows by having either the end or side closed (welded). When the patterns are superposed, no lines cross because this would create a weld all the way down the HX, precluding expansion. Some of the welds in one of the patterns could be omitted to increase the channel size, allowing a fluid that has higher viscosity to have similar head loss as the other side or allowing a fluid of lower density to move faster with similar head loss as the other side.

Figure 5. Laser welding pattern (in white): left: "bent" channels where the, e.g., hot water enters the lower left side and exits the upper right side; right: "straight" channels where the, e.g., cold water enters the top and exits the bottom.

Figure 6. Expanded HX manifolding.

*In this way, the small channels and manifolds are created with the accuracy of the laser, but then only macroscopic connections are required.*

*4.2 Manufacturing*

There are a variety of possible methods of expansion. The top and bottom could be pulled apart, but for non-stretching expansion, the device gets narrower. Therefore, perforated plates with air being drawn through them might be able to be used on top and bottom of the HX (however, this was unsuccessful for the prototype, despite lubrication). Another option is using a pressurized fluid to expand (used successfully for the prototype). A further expansion method possibility is gluing the tips of the bristles on a brush to the top and bottom of the HX, and pulling upward and downward. This would allow the HX to contract sideways, but removal of the brush might be difficult.

Though the construction of a prototype used a laser that traced out the weld, this results in an extremely expensive HX because of the laser time ($200/hr for the prototype). In order to make the manufacturing cost low, a continuous process is required. One possibility is a roll-to-roll process, where layers are added one at a time and welded on (see Figure 7). A mask would be used to turn a sheet of radiation into the desired welding pattern (the mask is represented stylistically in Figure 7 with the holes in the "roller mask"). The sheet of radiation could be produced by a laser beam that is expanded in one direction (welder on the right in Figure 7, the laser expander is not shown because it does not change the beam appearance in this perspective). Alternatively, the radiation from a heated filament could be collimated with a mirror and then focused with a lens (welder on the left in Figure 7).

The filament would only be practical in a limited number of circumstances. The laser and its energy cost compared to the material cost depends on a number of factors, but for a low-cost laser and one third of the area welded together, the following rough comparisons can be made. For most materials, the laser and its energy cost would be the same order of magnitude as material cost. For steel, the laser and its energy cost would be one order of magnitude larger than material cost. For Teflon, the laser and its energy cost would be one order of magnitude smaller than material cost. Finally for silver, the laser and its energy cost would be two orders of magnitude smaller than material cost.

Positioning accuracy of rolls of 0.01 mm is possible now [30], so 0.1 mm HX tubes would be feasible. After welding, the sheet would be cut up into pieces, expanded, and fixed in shape.

Figure 7. Mass production using filament and laser welding.

A cost optimization of HXs and the systems to which they are connected has been performed [31]. Though the conclusions vary depending on the heat capacity rate ratio and the application, the basic trade-off is that a lower cost per heat transfer ability can be used to lower the ineffectiveness (1-η)$_{opt}$, total HX expenditure E$_{HX}$ or some combination of the two. A typical scenario is where the difference is split, such that the optimal ineffectiveness (1-η)$_{opt}$ and total HX expenditure E$_{HX}$ scale according to:

$(1-\eta)_{opt} \propto E_{HX} \propto (P_{HT})^{-0.5}$ .                **Equation 12**

For instance, if the typical HX has a 30% ineffectiveness, and the cost per heat transfer ability for the mass-produced expanded microchannel HX is an order of magnitude less, the optimal ineffectiveness would be ~10%, and the total expenditure on the HX would be ~one third as much.

*4.3 Laser Welded Prototype*

Clear PP and polyethylene terephthalate (PET) sheets with reverse conduction laser welding (see section 2.2) were unsuccessful. Therefore, new type of laser welding was developed: using opaque polymer layers that absorb the laser and conduct the heat down through to the layer below – "forward conduction welding," – see Figure 8. This is similar to the reverse conduction laser welding discussed in section 2.2.

Figure 8. Forward conduction laser welding.

The intensity and exposure time of the laser have to be controlled very carefully, such that the weld does not weld three layers together (right weld on Figure 8).

The required patterns (see Figure 5) were programmed on a Leister Novolas WS laser welder with a 140 W maximum laser diode with 940 nm wavelength (6 W was used). The laser spot size was 1 mm and the speed was 50 mm/s, so this is an exposure time of 0.02 s, which would diffuse approximately 0.04 mm in LDPE, which agrees with observation (welding two layers together but not three). Black LDPE (garbage bags) with a wall thickness of 28 µm was successful.

We created a heat exchanger effectiveness model that used correlations to take into account axial conduction, heat loss to the environment, edge effect (the fact that peripheral channels have lower *NTU*) [32], and flow maldistribution (not all of the channels receiving the same flow) [33]. Assuming that maldistribution was only due to laser inaccuracy, the channel diameter variation would be approximately 5%, and the maximum predicted steady state effectiveness for liquid to liquid was 94%. For expansion, the friction between the plates and the HX was too great for this method to work with the prototype. Therefore, a pressurized fluid (air) was used to expand, which created ~2 mm rounded channels. Once the HX was expanded, a rigid frame was attached to maintain the shape.

**5. Measuring HX Effectiveness**

The effectiveness of the prototype was determined by measuring the incoming and outgoing water temperatures (see Figure 9). Rather than measuring the absolute temperatures, the temperature difference of the similar temperature flows (i.e. $T_{hi} - T_{co}$ and $T_{ho} - T_{ci}$) was measured to increase the accuracy [34]. With a given absolute accuracy, a higher initial temperature difference will yield a smaller percent error. Therefore, the low temperature was an ice bath, and the high temperature was as high as can be tolerated by the polymer (approximately 50°C for LDPE).

Figure 9. Experimental setup.

The error in the temperature measurement is composed of the error in the voltage sensor, switch, cold junction compensation, the thermocouple itself, logger resolution, and calibration. The overall error for an absolute temperature measurement assuming that all the errors are independent was calculated to be ~0.5°C. The percent error for each of the data points was calculated.

There is also the variation in the reservoir temperature over time. However, as this boundary condition changed, the quasi-steady-state effectiveness would still remain constant, so this would contribute a very small error. The variations in temperature across the cross section of the flow at the hot and cold outlet should not cause significant error.

The minimum flow rate of the HX prototype is only 0.01 mL/s, or 0.00015 gallons per minute. This low flow rate is difficult for conventional flow meters, but one simple accurate way of measuring the flow rate is the rate at which a graduated cylinder is filled. A needle valve was used to control the flow from the reservoirs to the graduated cylinders.

The total error in the flow rate was ~1% for most flow rates, primarily comprised of the graduated cylinder error (evaporation was negligible). However, for the flow rates above ~0.2 mL/s, the reading error became significant, so the total error was ~2%. The variation from the mean flow rate was ~1.5% due to variation in head.

Another experimental concern is bubbles in the HX. Since the solubility of gases in water decreases with increasing temperature, the input hot water would not produce bubbles upon cooling. However, the input cold water could produce bubbles upon heating. The highest measured output cold water temperature was 38°C. Therefore, the input cold water would need to have a "bubble point" (temperature at which bubbles are produced) of greater than 38°C. This was achieved by boiling water, putting it into canning jars, and refrigerating the jars. After running some trials, a sample of this mixed water was heated and the bubble point was 35°C. This was probably adequate for the typical channel because during the heating process to determine the bubble point, gases would diffuse into the water. However, some channels might have heated up to near the input hot water temperature of 50°C because of maldistribution, and this could have caused bubbles in the HX.

Another issue is the possibility that there were bubbles in the HX before the experiment started. To eliminate these, deaerated cold water was put in both reservoirs and run for approximately 8 hours, which should have been sufficient to absorb the bubbles in the channels with mean flow, but was probably inadequate for the channels with pinched flow.

The testing pressure conditions were maintained for several days before taking data to minimize polymer distortion error.

Precise control of the flow was not possible with the needle valve. Therefore, the flow was not perfectly balanced in all runs. For a HX with no heat leak, there is technically only one effectiveness value. However, for the rest of this paper, effectiveness will be used in the temperature sense, rather than heat flux:

$$\eta_h = \frac{T_{hi}-T_{ho}}{T_{hi}-T_{ci}},$$ **Equation 13**

and

$$\eta_c = \frac{T_{co}-T_{ci}}{T_{hi}-T_{ci}}.$$ **Equation 14**

The arithmetic mean of the above effectivenesses and the arithmetic mean of the *NTU* on the hot and the cold side were calculated and the data are plotted according to this *NTU* (see Figure 10). By using a temperature effectiveness, when the hot flow was considerably higher than the cold flow, the hot temperature effectiveness was considerably lower and the cold temperature effectiveness was considerably higher. Since the flow was unbalanced in one direction some of the times and the other direction other times, a graph of the hot and cold effectiveness would be erratic. This volatility was reduced by averaging the hot temperature effectiveness and the cold temperature effectiveness. This approximates the effectiveness that would be achieved if the flow were perfectly balanced.

The vertical error bars are 0.4% to 1.4% correspond to the uncertainty in temperature measurement, but error due to bubbles could have been significantly larger and is not quantified. On the logarithmic plot, it would not be possible to see the ~1.5% variation in *NTU* (resulting from the 1.5% variation in flow rate). This shows that the uncertainty in flow rate has a negligible influence on the uncertainty in effectiveness. The experimental effectiveness is considerably lower than the *NTU* model. The fall in effectiveness at intermediate *NTU* values is because the flow is significantly unbalanced, so averaging the effectivenesses does not produce the same result as balanced flow at the same average *NTU*.

Figure 10. Experimental and *NTU* model effectiveness as a function of average *NTU*.

## 6. Conclusions

Modeling indicates that high effectiveness can be achieved with polymer HXs. The new expanded microchannel HX has low material cost and should have low manufacturing cost. One new manufacturing method is forward conduction laser welding. The measured maximum effectiveness of the prototype with water-to-water heat transfer was 72%. The effectiveness was lower at lower *NTU* as with any heat exchanger, but the effectiveness also decreased at very high *NTU* because the losses to the environment became dominant at extremely low flow rates.

However, multiple serial stages of cross-flow HXs (low effectivenesses) can be used to approximate counter-flow (high) effectiveness [35]. Similarly, if multiple stages of the expanded microchannel HX are used, the penalty of maldistribution can be reduced. One way of thinking about the loss due to maldistribution is the entropy production that occurs upon mixing the individual streams that have different *NTU* and therefore different temperature. The entropy generation is proportional to the amount of heat transferred multiplied by the temperature difference. So if the HX has two stages, at each stage, it is mixing streams of the same mass with half the temperature difference, so half the heat is transferred at an average of half the temperature difference, so this is one fourth the entropy generation. Therefore, having two stages of mixing would have only half the overall entropy production. The resultant effectiveness is raised to the power of the reciprocal of the number of stages. For instance, for two stages, the effectiveness is raised to the one half power. Roughly, this cuts the ineffectiveness due to maldistribution in half.

Therefore, if three stages were used of the prototype microchannel expanded HX, 90% overall effectiveness would be achieved. Since the expanded microchannel HX has low material cost and there is the potential to reduce the manufacturing costs with mass production, multiple stages can be justified. Furthermore, because the reduction in pressure loss of laminar flow with lower velocity is accompanied by the maintenance of the heat transfer coefficient, the pressure loss of multiple stages can be kept low. Therefore, the microchannel expanded HX has the potential to be low-cost, high effectiveness, and low pressure loss. Therefore, the new low-cost HX has David C. Denkenberger, Michael J. Brandemuehl, Joshua M. Pearce, and John Zhai, "Expanded microchannel heat exchanger: design, fabrication and preliminary experimental test", *Proceedings of the Institution of Mechanical Engineers – Part A: Journal of Power and Energy*, **226**, 532-544 (2012). http://dx.doi.org/10.1177/0957650912442781the potential to save a significant amount of energy and thus reduce pollution at a lower initial cost. In addition, the new HX should make renewable energy more cost-effective. Furthermore, when applied to solar water pasteurization, it has the potential to save thousands of lives a day.

## 7. Future Work

Computerized Tomography scans and discretized computer modeling of the HX have been performed and will be submitted in a forthcoming article. This will show that the lower than expected effectiveness is due primarily to non-uniform channel dimensions causing flow maldistribution, and that alternate welding patterns have the potential to reduce the maldistribution problem.

*7.1 Laser Welding with Alternative Material*
The suitability of a material for forward conduction laser welding can roughly be characterized by the viable range divided by the maximum temperature rise (the boiling or decomposition temperature minus the ambient temperature). Values for candidate materials are shown in Table 1. Higher ambient temperatures increase the viable range fraction, but sufficient strength must be maintained for the continuous mass production process. Also, the cost increases due to the cost of maintaining a higher temperature and the components withstanding that higher temperature. Since the opaque LDPE at 20°C was successful (>64% range, as LDPE melts at a lower temperature than HDPE and should decompose at nearly the same temperature), all the metals could be made successful, though a significant increase in ambient temperature would be required for iron and titanium. Therefore, for high temperature applications that demand high corrosion resistance, silver could be used instead of stainless steel because only a small amount of material is required.

Table 1. Viable welding range fraction for different materials.

The pure or nearly pure forms of metals would be preferred to have great elongation (stretching before breaking) [36] for the expansion process (even non-stretching expansion requires bending of the material with accompanying plastic deformation). This means the tensile strength would be lower than the metal alloys, but it is still significantly higher than the polymer, so it can be used for high pressure and temperature applications. Furthermore, if the expansion process were performed at elevated temperature, even metal alloys behave as superplastics (greater than 100% elongation) [37], so higher strength metals could be used.

*7.2 Alternate joining methods and geometries*
"Inkjet" printing of adhesive should be tried, especially if laser welding does not work for polymer ceramic precursors. Adhesives that are metallic or ceramic-based may allow high-temperature, high-corrosion-resistance performance. Cross flow can be achieved with the new expanded HX with a welding pattern similar to the manifolds in the counter-flow arrangement. This would allow a large face area for the gas without a large manifold. Therefore, with materials including polymers, metals, and ceramics, and counter flow, parallel flow, and cross flow possible, the expanded HX should work for nearly all applications. One limitation is that the expanded HX cannot be disassembled for physical cleaning, so in applications where there is high fouling and chemical cleaning is insufficient, the expanded HX would not work.

**Acknowledgements**
Valuable feedback was received from Ray Radebaugh and Moncef Krarti. Experimental assistance was provided by Jill Stone and Andrew Bour.**Declaration of Conflicting Interests**
Support for this project was received from the American Society of Heating, Refrigeration, and Air Conditioning Engineers Graduate Research Fellowship and the University of Colorado at Boulder Technology Transfer Office Proof of Concept Grant. These funding sources had negligible influence on the research and publication process.
Dr. Denkenberger is the president of Denkenberger Inventing and Consulting, LLC which has licensed the microchannel expanded heat exchanger from the University of Colorado and plans to commercialize the technology.

**Appendix: Notation**

| Symbol | Units | Explanation |
|---|---|---|
| A | $m^2$ | Heat transfer area |
| C | - | Heat capacity rate ratio |
| $C_c$ | kJ/(s*K) | Cold heat capacity rate |
| $C_h$ | kJ/(s*K) | Hot heat capacity rate |
| $C_{min}$ | kJ/(s*K) | Minimum heat capacity rate |
| $C_{max}$ | kJ/(s*K) | Maximum heat capacity rate |
| $C_p$ | J/(kgK) | Specific heat at constant pressure |
| $E_{HX}$ | $ | Expenditure on HX |
| h | W/($m^2$K) | Heat transfer coefficient |
| HDPE | - | High density polyethylene |
| HX | - | Heat exchanger |
| k | W/(mK) | Thermal conductivity |
| LDPE | - | Low density polyethylene |
| $\dot{m}$ | kg/s | Mass flow rate |
| NTU | - | Number of transfer units |
| Nu | - | Nusselt number |
| $P_{HT}$ | $/(W/K) | Price per heat transfer ability |
| $P_V$ | $/$m^3$ | Price of HX material per volume |
| PET | | Polyethylene terephthalate |
| PP | - | Polypropylene |
| PS | - | Polystyrene |
| $\dot{q}$ | W | Heat transfer rate |
| t | m | Wall thickness |
| T | °C | Temperature |
| U | W/($m^2$K) | Overall heat transfer coefficient |
| **Greek** | | |
| η | - | Effectiveness |
| (1-η) | - | Ineffectiveness |
| **Subscripts** | | |
| c | - | Cold |
| h | - | Hot |
| HT | - | Heat transfer |
| i | - | In |
| max | - | Maximum |
| o | - | Out |
| opt | - | Optimum |
| t | - | Total |
| w | - | Wall |

David C. Denkenberger, Michael J. Brandemuehl, Joshua M. Pearce, and John Zhai, "Expanded microchannel heat exchanger: design, fabrication and preliminary experimental test", *Proceedings of the Institution of Mechanical Engineers – Part A: Journal of Power and Energy*, **226**, 532-544 (2012).  http://dx.doi.org/10.1177/0957650912442781**References**

1 Zaheed, L. and Jachuck, R.J.J. Review of polymer compact heat exchangers, with special emphasis on a polymer film unit. Appl. Therm. Eng. 2004, 24 (16), 2323-2358.
2 Greenblatt, J.B. Succar, S., Denkenberger, D., Williams, R. H., Socolow, R. H. Baseload wind energy: modeling the competition between gas turbines and compressed air energy storage for supplemental generation. Energy Policy, 2007, 35, 1474-1492.
3 United Nations Development Programme. Human development report 2001: Making new technologies work for human development. Oxford University Press, Oxford, 2001.
4 Denkenberger, D.C., Pearce, J.M. Compound parabolic concentrators for solar water heat pasteurization: numerical simulation. Proc. Sol. Cook. Int. Conf., 2006, July, Granada, Spain.
5 McQuiston, F.C., Parker, J.D., Spitler, J.D. Heating, ventilating, and air conditioning analysis and design. Sixth Edition. John Wiley & Sons, Inc. USA, 2005, 623 pages.
6 Doty, F.D, Hosford, G., Jones, J.D., and Spitzmesser, J.D., A laminar-flow heat exchanger. Proceedings of the Intersociety Energy Conversion Engineering Conference, 1990.
7 Pearce, J.M., Denkenberger, D.C., Zielonka, H. Energy conservation measures as investments, in: Spadoni, Giacomo. editor, Energy Conservation: New Research, Nova Science Publishers: Hauppauge, NY 2009, pp. 67-85.  ISBN: 978-1-60692-231-6.
8 Ashman, S., Kandlikar, S. A review of manufacturing processes for microchannel heat exchanger fabrication in: Proc. Fourth Int. Conf. on Nanochannels, Microchannels and Minichannels, June, Limerick, Ireland, 2006.
9 Lowenstein, A. A zero carryover liquid-desiccant air conditioner for solar applications. ASME/SOLAR06, July, Denver, CO, USA, 2006.
10 Veltkamp, W.B. Heat exchanger. United States Patent 5,725,051, March 10, 1998.
11 Carman, B.G., Kapat, J.S. Chow, L.C., An, L. Impact of a ceramic microchannel heat exchanger on a micro turbine. Proc. ASME Turbo Expo 2002, June, Amsterdam, The Netherlands, 2002, 1053-1060.
12 Perry, C.R., Dietz, L.H, Shannon, R.L. Heat exchange apparatus having thin film flexible sheets. United States Patent 4,411,310, October 25, 1983.
13 Ramm-schmidt, L. Apparatus for heat transfer between gas flows. United States Patent 6,758,261, July 6, 2004.
14 Bachmann, F.G., Russek, U.A. Laser welding of polymers using high power diode lasers. Photonics West 2002 Conf., June, San Jose, CA, USA, 2002.
15 Garst, S., Schuenemanna, M., Solomon, M., Atkinb, M., Harveyac, E. Fabrication of multilayered microfluidic 3D polymer packages. 2005 Electronic Compon. and Technology Conf.
16 Doalman, J.D.B. The mystery of electronic water treatment unveiled. Industrial Technology News, November 25, 2005.
17 Shah, R. K., Focke, W.W. Plate heat exchangers and their design theory. in: Shah, R. K.,  Subbarao, E. C., Mashelkar, R. A. (eds.), Heat Transfer Equipment Design, Hemisphere Publishing, Washington D.C., 1988.
18 Goddard, J.B. Abrasion resistance of piping systems. ADS, 1994.
19 Powell, M.R., Hong, S.H., Paulsen, S. Recirculating thermocatalytic air purifier for collective protection. NBC Def. Collective Protection Conf., 2002.
20 Ahuja, V., Green, R. Carbon dioxide removal from air for alkaline fuel cells operating with liquid hydrogen: A synergistic advantage. Int. J. of Hydrogen Energy. 1998, 23 (2), 131-137.
21 Wilson, D.G. Wilson turbopower's David Gordon Wilson presents seminal scientific paper at international turbine congress; peer-reviewed paper outlines the theory and design of Wilson turbopower's new revolutionary heat exchanger. Business Wire, May 15, 2006.
22 Creare Heat exchangers. http://www.creare.com/services/fluid/heat_ex.html accessed March 23, 2007.
23 Ganzevles, F.L.A. Drainage and condensate heat resistance in dropwise condensation of multicomponent mixtures in a plastic plate heat exchanger. PhD. Thesis, 2002, Technishe Universiteit Eindhoven.
24 Hart, G.K., Lee, C-o, Latour, S.R. Development of plastic heat exchangers for ocean thermal energy conversion. 141 pages, A092503, Jan 1979.
25 Greenbox. http://www.greenbox.uk.com/ accessed April 1, 2008.
10

## Figures and Captions

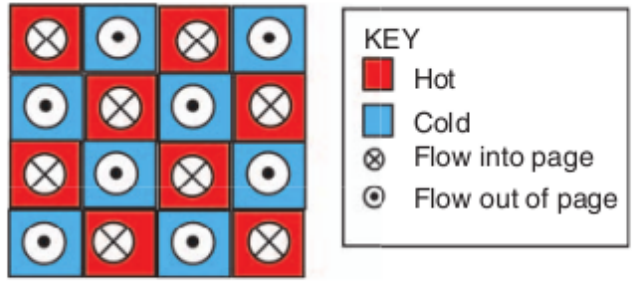

**Fig. 1** Cross-sectional end view of the extruded 'chess-board' pattern, e.g. [11]

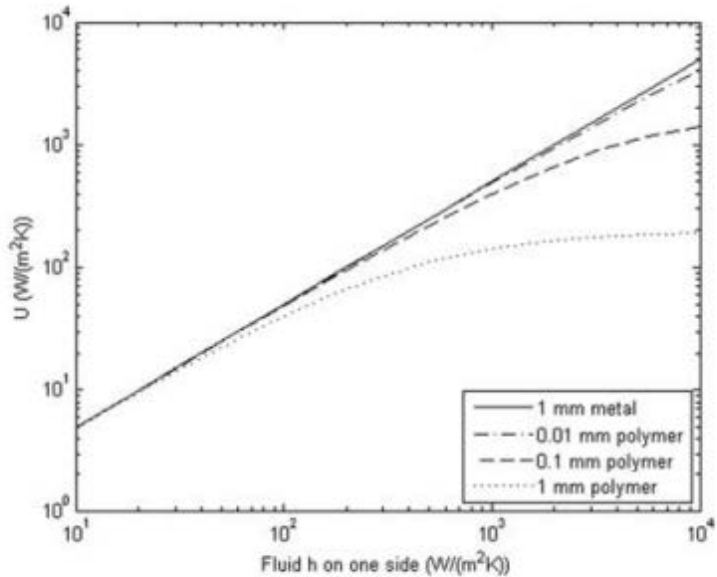

**Fig. 2** Overall heat transfer coefficient with metal and different polymer thicknesses (mm)

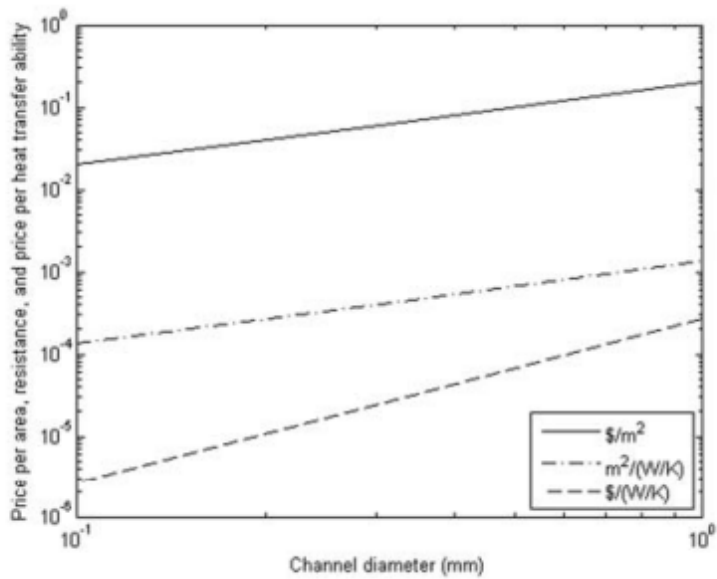

**Fig. 3** Price per area ($/m²), thermal resistance (m²/(W/K)), and price per heat transfer ability ($/(W/K)) as a function of channel diameter (mm) with wall thickness 10 per cent of diameter

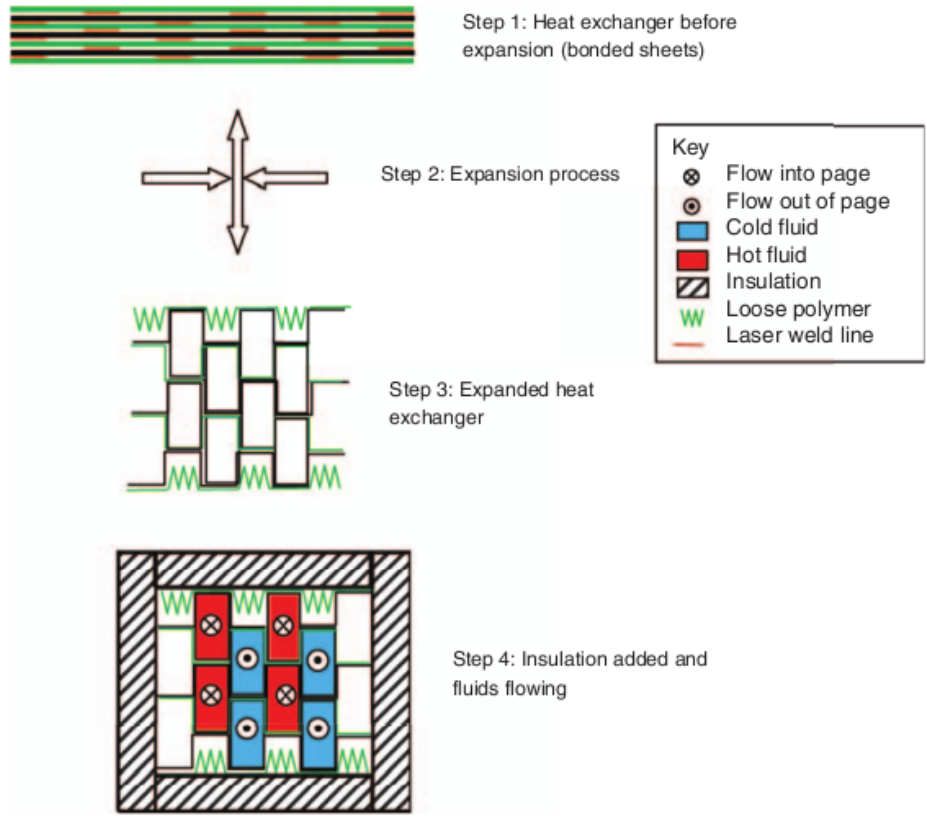

**Fig. 4** Cross-sectional view of the expansion process for the HX core
Note: The start is the top and the bottom is expanded (with insulation put on the outside)

David C. Denkenberger, Michael J. Brandemuehl, Joshua M. Pearce, and John Zhai, "Expanded microchannel heat exchanger: design, fabrication and preliminary experimental test", *Proceedings of the Institution of Mechanical Engineers – Part A: Journal of Power and Energy*, **226**, 532-544 (2012).  http://dx.doi.org/10.1177/0957650912442781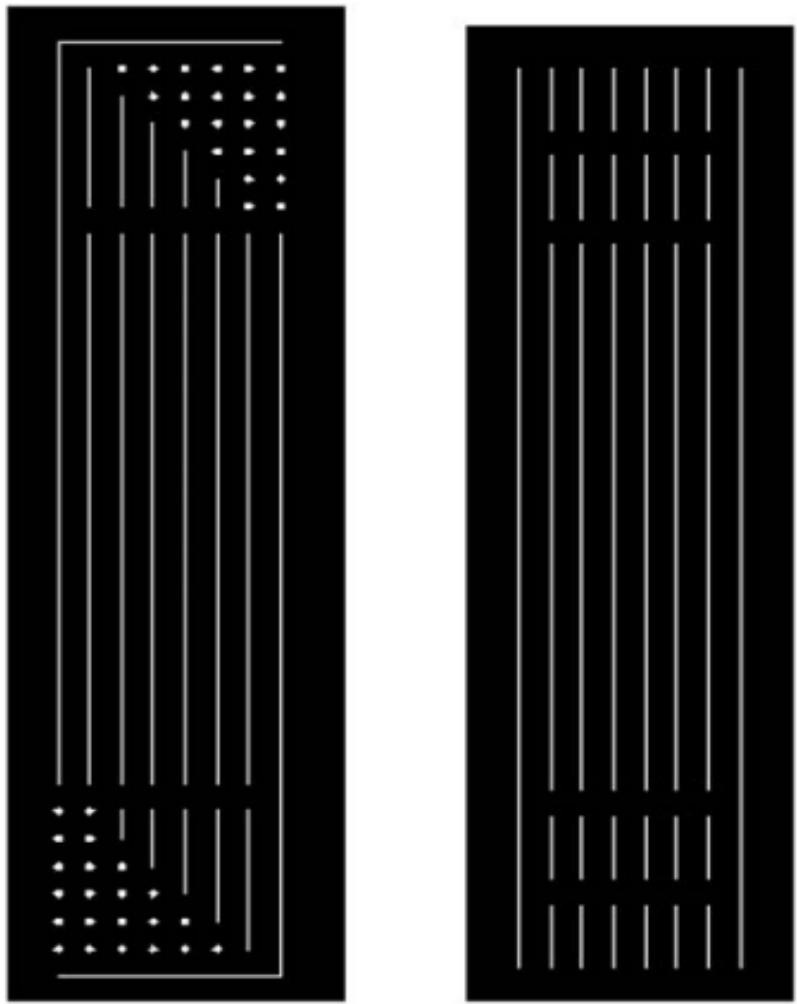

**Fig. 5** Laser welding pattern (in white): left: 'bent' channels where the, e.g. hot water enters the lower left side and exits the upper right side; right: 'straight' channels where the, e.g. cold water enters the top and exits the bottom

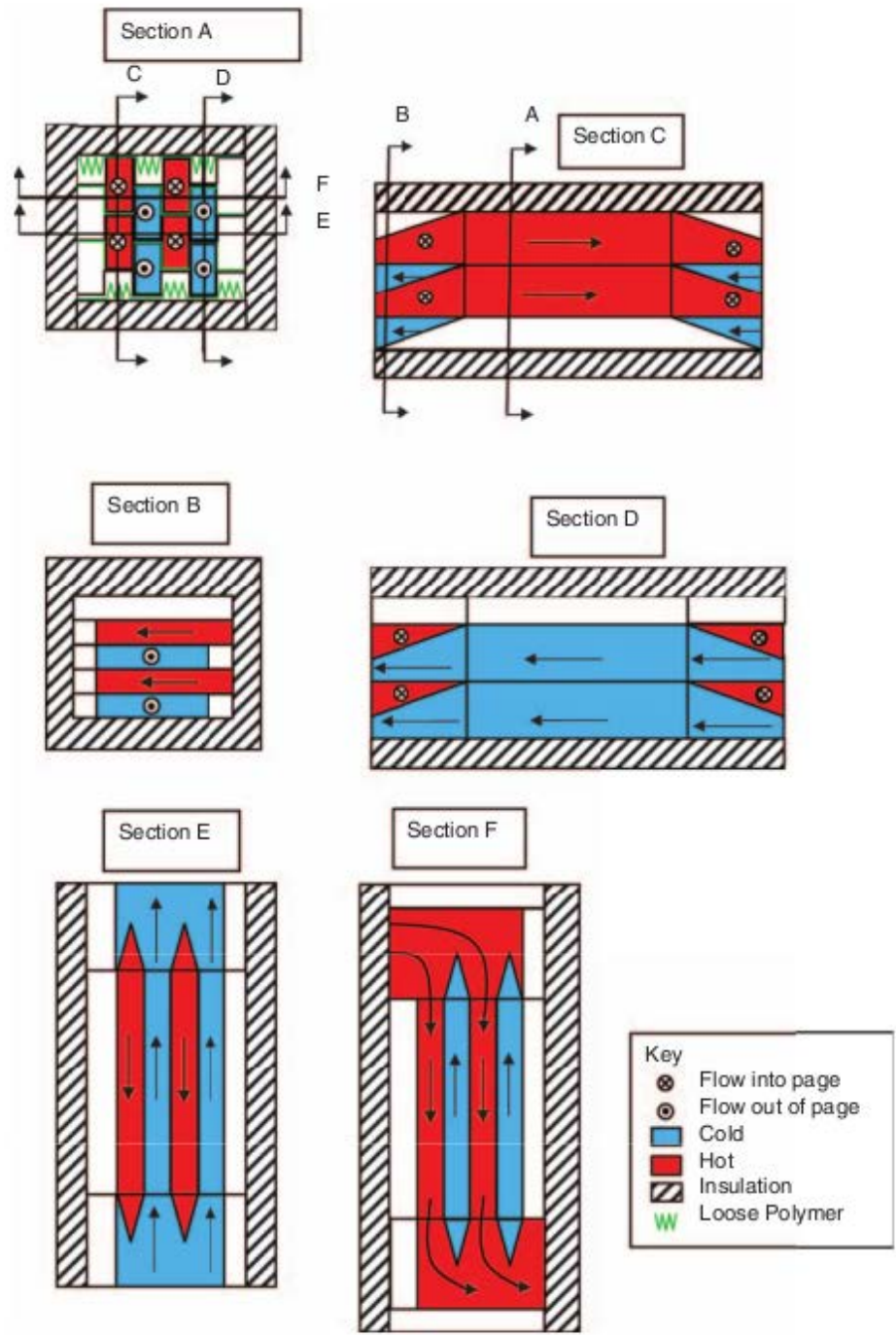

**Fig. 6** Expanded HX manifolding

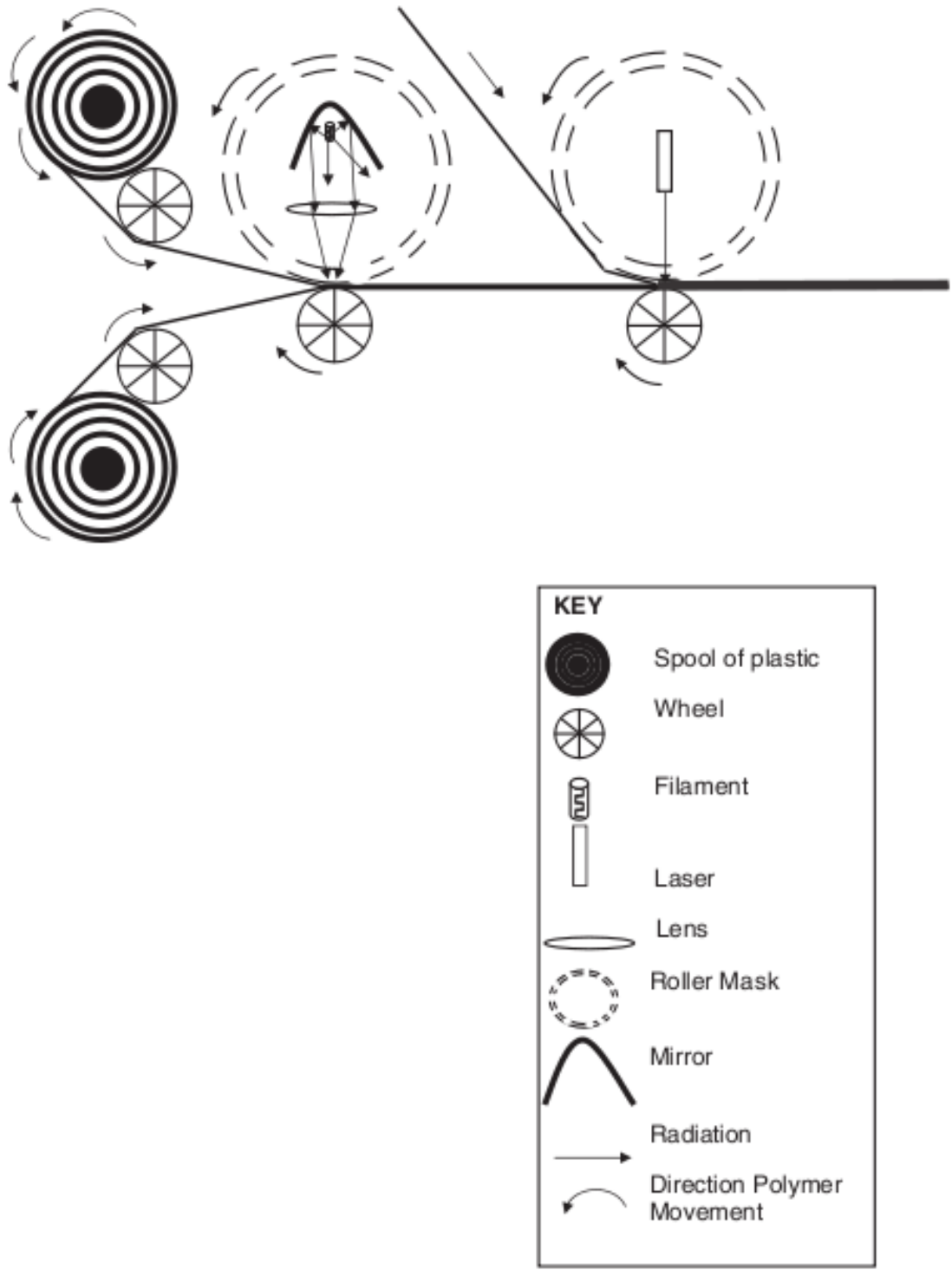

Fig. 7  Mass production using filament and laser welding

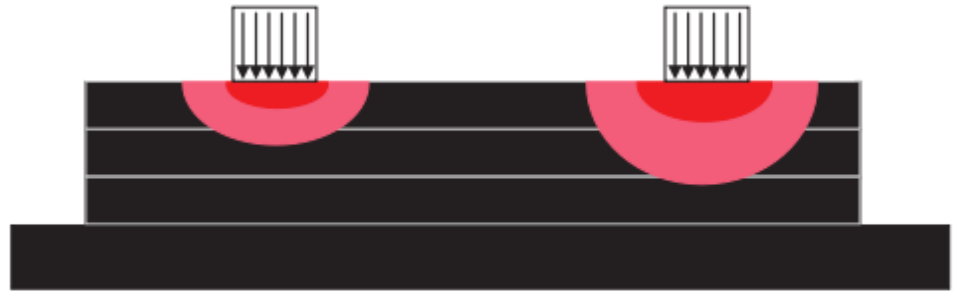

**Fig. 8** Forward conduction laser welding: three black polymer layers rest on a black metal plate, arrows represent the laser, and reds represent the melt area

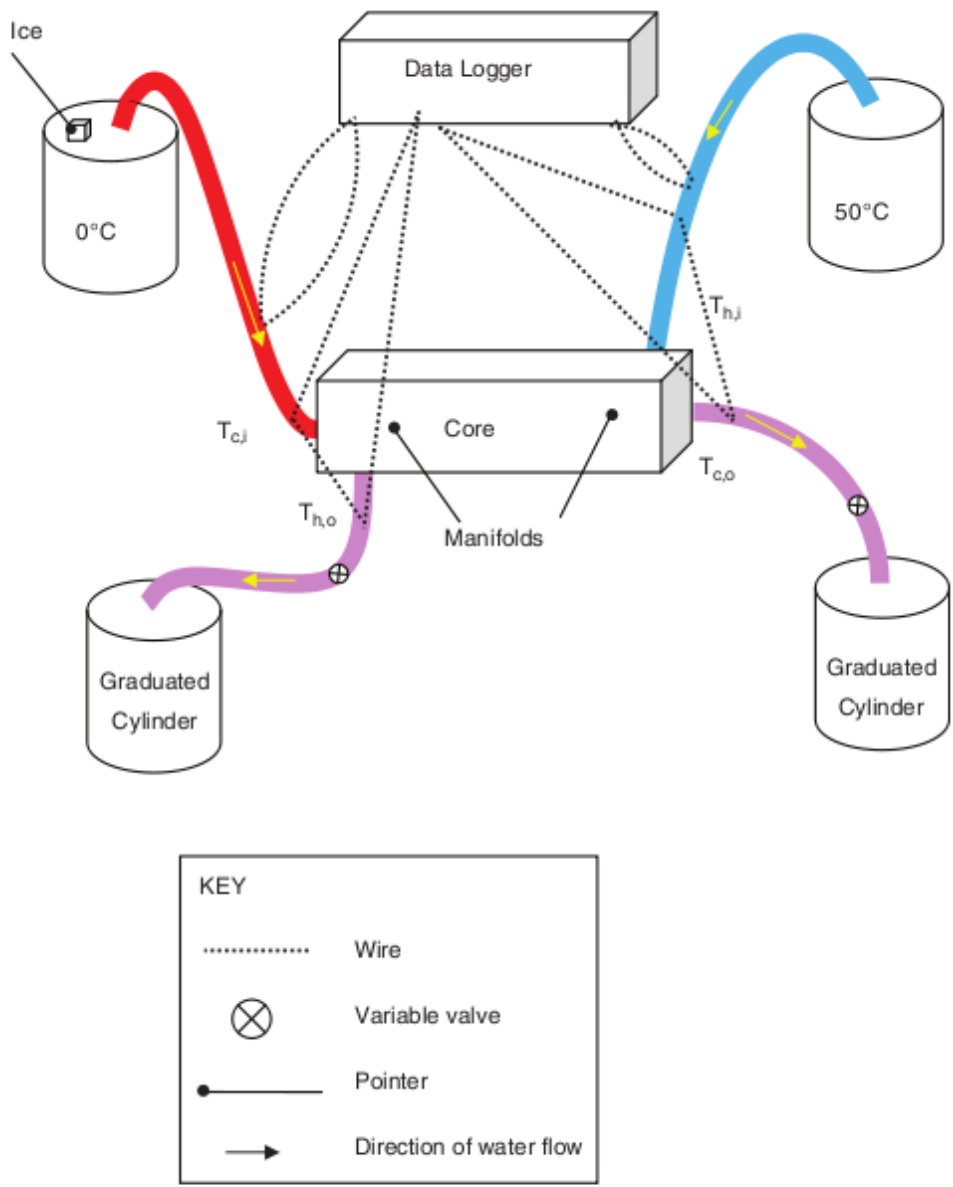

Fig. 9   Experimental setup

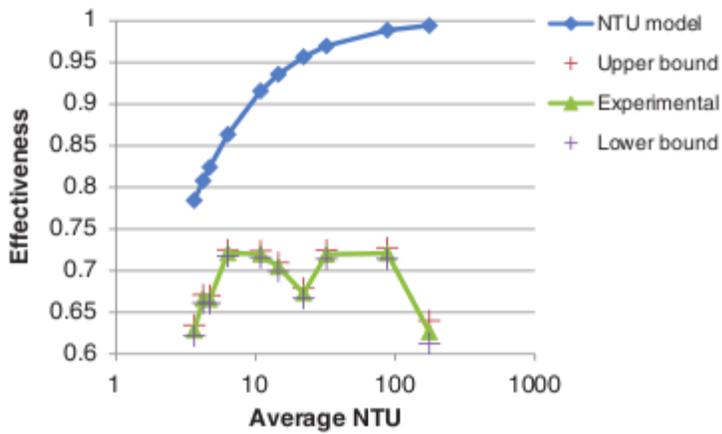

**Fig. 10** Experimental and NTU model effectiveness as a function of average NTU